\documentclass{article}[11pt]
\usepackage{amsfonts}
\usepackage{CJK,amsmath,indentfirst}
\usepackage{graphics}
\usepackage{graphicx}
 \small

\newcommand{\be}{\begin{equation}}
\newcommand{\nd}{\end{equation}}
\newcommand{\bea}{\begin{eqnarray}}
\newcommand{\nda}{\end{eqnarray}}
\newcommand{\p}{\partial}
\newcommand{\nn}{\nonumber}

\newcommand{\fl}{\hspace*{-45pt}}

\begin{document}

\title{Option Pricing with Lie Symmetry Analysis and Similarity
Reduction Method}

\author{Wenqing Bao$^{1,2}$ \qquad Chun-Li Chen$^{1}$\footnote{corresponding author. Email:
clchen@sjtu.edu.cn.}\qquad Jin E. Zhang$^3$\\
\footnotesize $^{1}$\it Department of Mathematics,Shanghai Jiao Tong
University, Shanghai, 200240, P. R. China\\
\footnotesize $^{2}$\it Antai College of Economics and Management
Shanghai Jiao Tong University, Shanghai, 200240, P. R. China\\
\footnotesize $^{3}$\it Department of Accountancy and Finance, Otago
Business School, University of Otago, Dunedin 9054, New Zealand}

\date{}
\maketitle
\parindent=0pt

\begin{center} {\bf Abstract}  \end{center}

With some transformations, we convert the problem of option pricing
under state-dependent volatility into an initial value problem of
the Fokker-Planck equation with a certain potential. By using the
Lie symmetry analysis and similarity reduction method, we are able
to reduce the dimensions of the partial differential equation and
find some of its particular solutions of the equation. A few case
studies demonstrate that our new method can be used to produce
analytical option pricing formulas for certain volatility functions.

\vskip 1cm


\vskip 0.5cm


\section{Introduction}

The landmark works of Black and Scholes (1973) and Merton (1973)
have created a new field in quantitative finance. In the
Black-Scholes/Merton framework, the price of an underlying asset is
often modeled as a diffusion process. With a no-arbitrage argument,
the price of a derivative contract written on the asset can be
determined by solving an initial boundary value problem of a linear
partial differential equation (PDE). In the classical Black-Scholes
model, the volatility of the underlying asset, $\sigma$ is assumed
to be constant. In order to explain the empirical phenomenon of
implied volatility smirk, see e.g., Zhang and Xiang (2008),
researchers propose to use volatilities defined by deterministic
functions of the underlying asset price and time. The corresponding
PDE is often called the generalized Black-Scholes equation.
Analytical formulas of the problem for the general case is not
available. However, the problem of reducibility and solvability of
the generalized Black-Scholes equation has been studied by Carr {\it
et al.} (1999, 2002, 2006), Bouchouev (1998), Li and Zhang (2004),
and Zhang and Li (2012). Haven (2005) suggestions a solution
technique for obtaining analytical solutions to the generalized
Black-Scholes equation via an adiabatic approximation to the
Schr\"odinger PDE.

In 1891, a famous mathematician, Sophus Lie, pointed out that, if an
ordinary differential equation (ODE) is invariant under a
one-parameter Lie group of transformations,  the order of the ODE
can be reduced constructively. The method of finding similarity
reductions of a given PDE by using the Lie group method of
infinitesimal transformation (sometimes called the method of
group-invariant solutions) was originally developed by Lie (1891),
see Olver (1993) for the recent developments. Bluman and Cole (1969)
proposed a generalization of Lie's method which is called the
nonclassical method of group-invariant solutions. The method was
further generalized by Olver and Rosenau (1986). A common feature of
these methods is to determine Lie point transformations of a given
PDE, i.e., transformations that depend only on the independent and
dependent variables, see equation (\ref{LieTrans}). After that, Lie
group analysis was widely applied in solving differential equations
in fluid mechanics and quantum mechanics.
Nowadays Lie symmetry software packages are widely used in solving PDEs.
Reviews and comparative studies of some of the earlier computer algebra packages
have been carried out by Hereman (1997) and Butcher {\it et al.} (2003).
More recently, Rocha Filho and Figueiredo (2011) presented the new MAPLE package SADE
for the determination of symmetries and related properties of systems of differential
equations. Vu {\it et al.} (2012) presented the new MAPLE symmetry package DESOLVII,
an upgrade of DESOLV, which included the functionality to determine higher classical
symmetries for both ordinary and partial differential equations. However, currently,
in the situation that coefficient functions contain arbitrary functions,
software packages for symmetry analysis cannot handle.

In fluid mechanics and quantum mechanics, the determination of the
symmetry group of Fokker-Planck equations has a long history. Finkel
(1999) completely classified the symmetries of the Fokker-Planck
equation and constructed group-invariant solutions for a physically
interesting family of Fokker-Planck equations in the case of two
spatial dimensions, namely
\begin{equation}\label{eqFP}
    u_t(x,y,t)-\frac{1}{2}\Delta u(x,y,t)+M(x,y,t)u=0,
\end{equation}
where $u$ is a dependent variable and $M(x,y,t)$ is a potential
function. Building on Finkel's result, Laurence and Wang (2005)
found some closed-form fundamental solutions for a special family of
Fokker-Planck equations. They showed how these results can be
applied in finance to yield exact solutions for special affine and
quadratic two factor term structure models. 
In this paper, we not only show how to
generate a series of new solutions with a given solution by using
the last set of equations in Appendix A, but also perform similarity
reductions of different cases.

In quantitative finance, Lo and Hui (2001) presented Lie-algebraic
method for the valuation of financial derivatives with
time-dependent parameters based upon the Wei-Norman theorem. Lo and
Hui (2002) extended their Lie-algebraic approach for the valuation
of multi-asset financial derivatives in a lognormal framework with
time-dependent parameters (drift, standard-deviation, correlation),
involving also stochastic short-term interest rates. Lo and Hui
(2006) proposed also a Lie-algebraic model for pricing more complex
derivatives like moving barrier options with time-dependent
parameters in a CEV framework. The difference between Lo and Hui's
Lie-algebraic approach and our Lie symmetry approach is as follows.
In our Lie symmetry approach, we obtain similarity reduction by
using a one-parameter invariant group of partial differential
equations. Lie algebras are by-products after we obtain the vector
fields in equation (\eqref{symmetries}). However, Lo and Hui (2001)
start from a Lie algebra, which is elevated to a group via an
exponential mapping. Carr, Laurence and Wang (2006) performed the
classification of driftless time and state dependent diffusions that
are integrable in closed form via Lie's equivalence transformations.
However, the Lie symmetry analysis and similarity reduction of the
generalized Black-Scholes equation (with general volatility
function) are not available yet.

In this paper, we try to solve the problem of option pricing based
on the theory of the Fokker-Planck equation. The $2-$dimensional
generalized Black-Scholes equation, arising from option pricing, can
be transformed into the $2+1-$dimensional Fokker-Planck equation. We
demonstrate how to apply Lie symmetry analysis and similarity
reduction to solve the option pricing problem for volatility as a
function of underlying asset price.

Compared with Lo and Hui's approach, our Lie symmetry approach is
more systematic. The main purpose of this paper is to demonstrate
the methodology by using a state-dependent volatility, $\sigma(S)$.
If the volatility is state- and time-dependent, $\sigma(S,t)$, then
the potential function, $M$, in equation (9) is also a function of
time, i.e., $M(x,y) \to M(x,y,t)$. Our approach can be used to
handle the case in principle as shown by equation (12). The
application to the case of state-and-time-separable volatility,
$\sigma(S,t) = \sigma_1(S) \sigma_2(t)$, will be reported in a
subsequent research\footnote{Lo and Hui's (2001, 2006)
time-dependent CEV, $\sigma(t) S^{\beta/2}$, is a special case of
state-and-time-separable volatility.}.  For the case of only one CEV
process, the parameter $\alpha$ in our Section 5.2 can take any
non-negative value, while Lo and Hui (2001, 2006) focus on $0 \le
\beta < 2$, which is equivalent to our $0 \le \alpha < 1$.

This paper is organized as follows. Section 2 discusses how to
transform a typical option pricing problem into the Fokker-Planck
equation like \eqref{eqFP}. Section 3 applies the Lie symmetry
analysis to the equation. Section 4 presents the similarity
reductions of different cases. Section 5 provides a few exact
solutions of both $2-$dimensional and $1-$dimensional generalized
Black-Scholes equation. Finally, section 6 concludes.

\section{Typical option pricing problem}

In the Black-Scholes (1973)/Merton's (1973) framework, the prices of
two stocks, $S_1$, $S_2$, are modeled by two pure diffusion
processes
\begin{equation}\label{dp}
    dS_i = \mu_i S_i dt + \tilde \sigma_i(S_i) S_i dB_i, \quad \quad
    (i=1,2),
\end{equation}
where $\mu_i$ is the drift, $\tilde \sigma_i(S_i)$ is the volatility
of the stock $i$, and $B_i$ ($i=1,2$) are standard Brownian motions.
The correlation coefficient between $B_1$ and $B_2$ is $\rho$. The
correlation makes it harder to convert the equation to $2+1$
dimensional Fokker-Planck equation like \eqref{eqFP}. In this paper,
we only consider the case where the volatility, $\sigma$, is a
deterministic function of the stock price, and leave the general
case of time dependence for future research. Standard no-arbitrage
theory shows that the price of a European style option,
$c(S_1,S_2,t)$, satisfies the following generalized Black-Scholes
equation
\begin{equation}
  \frac{\partial c}{\partial t} +\frac{1}{2}\sigma_1^2(S_1)
  \frac{\partial^2c}{\partial S_1^2}+
  \rho\sigma_1(S_1)\sigma_2(S_2)\frac{\partial^2c}{\partial S_1 \partial S_2}
  +\frac{1}{2}\sigma_2^2(S_2)\frac{\partial^2c}{\partial S_2^2}
  +rS_1\frac{\partial c}{\partial S_1}+rS_2\frac{\partial c}{\partial
  S_2}-rc=0,
  \label{eq0} \end{equation}\begin{equation}
  c(S_1,S_2,T)=C(S_1,S_2), \label{eq1}
\end{equation}
where correlation coefficient $\rho$ and interest rate $r$ are
assumed to be constant; $C(S_1,S_2)$ is the payoff function of the
option on the maturity date ($t=T$). For brevity, we have used
$\tilde \sigma_i(S_i)S_i = \sigma_i(S_i)$.

In the general case, analytical formulae of the problem \eqref{eq0}
and \eqref{eq1} cannot be obtained. Practitioners rely on numerical
methods such as finite difference, binomial trees, or Monte Carlo
simulation. However, in the way of reducibility and solvability of
the $1+1-$dimensional generalized Black-Scholes equation, Li and
Zhang (2004) determined the boundary condition and the nature of the
eigenvalues and eigenfunctions with Weyl-Titchmarsh theory. The
solution can be written analytically in a Stieltjes integral. Zhang
and Li (2012) provide a systematic way of finding the volatility
function, $\sigma(S)$, for a given solvable potential function.

Analytical solutions for the generalized Black-Scholes equation are
of paramount importance to practitioners as they allow a better
qualitative understanding of the solution behavior. More
significantly, volatility functions are typically fitted to market
data in empirical research. Parametric volatility models that
produce analytical solutions are in very high demand.

For certain volatility functions, e.g., a volatility being a
quadratic function of asset price studied by Z\"uhlsdorff (2001),
the generalized Black-Scholes equation can be transformed into the
standard heat equation, which, in turn, can be solved analytically.
Even for the case where the problem cannot be reduced to the
standard heat equation, it is still possible to solve the problem
analytically for some particular volatility functions. This paper
pushes further along this direction.

With the following transformation
\begin{equation}\label{trans1}
\begin{cases}
   \displaystyle x = \sqrt{\frac{2}{1+\rho}} \bigg(\int_0^{S_2}\frac{1}
   {\sigma_2(S)}dS + \int_0^{S_1}\frac{1}{\sigma_1(S)}dS \bigg),\\
   \displaystyle y = \sqrt{\frac{2}{1-\rho}} \bigg(\int_0^{S_2}\frac{1}
   {\sigma_2(S)}dS - \int_0^{S_1}\frac{1}{\sigma_1(S)}dS \bigg),
\end{cases}
\end{equation}
equations \eqref{eq0} and \eqref{eq1} become
\begin{equation}
    \frac{\partial c}{\partial t}+\frac{1}{2}\bigg(\frac{\partial^2c}
    {\partial x^2} +\frac{\partial^2c}{\partial y^2}\bigg)
    +Q_1\frac{\partial c}{\partial x} + Q_2\frac{\partial c}{\partial y}-rc=0,
    \label{eq2}\end{equation}\begin{equation}
    c(x,y,T)=C(x,y), \label{eq3}
\end{equation}
where
\begin{equation}
\begin{cases}
   \displaystyle Q_1 = \sqrt{\frac{2}{1+\rho}}\bigg(\frac{rS_1}{\sigma_1}
   +\frac{rS_2}{\sigma_2}\bigg)
   -\frac{2}{1+\rho}\bigg(\frac{\sigma_{1x}}{\sigma_1}
   +\frac{\sigma_{2x}}{\sigma_2}\bigg)
   +\frac{2}{\sqrt{1-\rho^2}}\bigg(\frac{\sigma_{1y}}{\sigma_1}
   -\frac{\sigma_{2y}}{\sigma_2}\bigg),\\
   \displaystyle Q_2 = \sqrt{\frac{2}{1-\rho}}\bigg(\frac{rS_1}{\sigma_1}
   +\frac{rS_2}{\sigma_2}\bigg)
   -\frac{2}{\sqrt{1-\rho^2}}\bigg(\frac{\sigma_{1x}}{\sigma_1}
   -\frac{\sigma_{2x}}{\sigma_2}\bigg)
   -\frac{2}{1-\rho}\bigg(\frac{\sigma_{1y}}{\sigma_1}
   +\frac{\sigma_{2y}}{\sigma_2}\bigg),\nonumber
\end{cases}
\end{equation}
$\sigma_{ix}$ and $\sigma_{iy}$ ($i=1, 2$) stand for partial
derivative of $\sigma_i$ with respect to $x$ and $y$ respectively,
$S_i$ is a function of $x,y$, which can be solved by equation
\eqref{trans1}. Consequently, $\sigma_i(S_i)$ converts to
$\sigma_i(x,y)$. For brevity, we have replaced $\sigma_i(x,y)$ by
$\sigma_i$.

We introduce the following transformation
\begin{equation}\label{trans2}
    c(x,y,t) = e^{\omega(x,y)-r\tau} u(x,y,\tau), \quad \quad \tau = T-t,
\end{equation}
where $\displaystyle \nabla \omega(x,y) = -(Q_1,\,Q_2)$. Here we
need following compatibility condition\footnote{If the condition is
not satisfied, the generalized Black-Scholes equation (\ref{eq0})
will be converted into a general case of Fokker-Planck equation,
instead of the irrotational case studied in this paper. It is
possible to study the solution of the general case of Fokker-Planck
equation by using Lie symmetry approach. The result will be reported
in a subsequent research.}
\be \frac{\p Q_2}{\p x} = \frac{\p Q_1}{\p y}. \nn \nd
By a simple calculation, equations \eqref{eq2} and \eqref{eq3}
become $2+1-$dimensional Fokker-Planck equation like \eqref{eqFP}:
\begin{equation}
    \frac{\partial u}{\partial \tau}-\frac{1}{2}\bigg(\frac{\partial^2u}{\partial x^2}
    +\frac{\partial^2u}{\partial y^2}\bigg)+M(x,y)u=0, \label{eq4}
\end{equation}
\begin{equation}
    u(x,y,\tau)|_{\tau=0}=u_0(x,y), \label{eq5}
\end{equation}
where the coefficient $M(x,y)$, regarded as a potential function,
reads
\begin{equation}\label{M}
    M(x,y) = \frac{1}{2}\bigg(\frac{\partial Q_1}{\partial x} +\frac{\partial Q_2}
    {\partial y}+Q_1^2+Q_2^2\bigg).
\end{equation}

Similarly, we can convert the $1-$dimensional generalized
Black-Scholes equation to $1+1-$dimensional Fokker-Planck equation.
Carr, Laurence and Wang (2006) exploit a remarkable intertwining
with the inhomogeneous Burger's equation in the time dependent and
state dependent one dimensional case via point transformations. By
using the separating variable method, Li and Zhang (2004), and Zhang
and Li (2012) transformed the option pricing problem into a
Schr\"{o}dinger equation which is similar to the $1+1-$dimensional
Fokker-Planck equation studied here\footnote{Li and Zhang (2004),
and Zhang and Li (2012) study the pricing of European options
written on a single asset, while we are studying case of two assets.
Their transformation is similar to a single-asset case of ours here
without the drift of risk-free rate.}.

\section{Lie point symmetries}

We now perform Lie symmetry analysis for the $2+1-$dimensional
Fokker-Planck equation. Let us consider a $2+1-$dimensional equation
\begin{equation}\label{sysfu}
    F(u)=u_t(x,y,t)-\frac{1}{2}\Delta u(x,y,t)+M(x,y,t)u.
\end{equation}
and a one-parameter Lie group of infinitesimal
transformation\footnote{More explanation of the treatment and the
meaning of the variables can be found in Chapters 2 and 3 of Olver's
(1993) book.}
\begin{eqnarray}
   & t \rightarrow t + \epsilon T(x,y,t,u),\nonumber\\
   & x \rightarrow x + \epsilon X(x,y,t,u),\nonumber\\
   & y \rightarrow y + \epsilon Y(x,y,t,u),\nonumber\\
   & u \rightarrow u + \epsilon U(x,y,t,u).\label{LieTrans}
\end{eqnarray}
With a small parameter $\epsilon \ll 1$, the vector field associated
with the group of transformations \eqref{LieTrans} can be written as
\begin{equation}\label{vector1}
   \underline{u} = T\frac{\partial}{\partial t}+X\frac{\partial}{\partial x}
                     +Y\frac{\partial}{\partial y}+U\frac{\partial}{\partial u},
\end{equation}
or equivalently in the symmetry form
\begin{equation}\label{sym1}
 \sigma = U-Tu_t-Xu_x-Yu_y.
\end{equation}
We wish to determine all possible coefficient functions $X,\,Y,\,T$
and $U$, so that the corresponding one-parameter group is a symmetry
group of the Fokker-Planck equation. The symmetry equation, i.e. the
corresponding infinitesimal criterion becomes
\begin{equation}\label{syseq2}
   \frac{\partial}{\partial \epsilon} \,F\big(u+\epsilon\sigma \big)
   \bigg|_{\begin{subarray}{1}\epsilon=0 \\ F(u)=0\end{subarray}}=0.
\end{equation}
Based on \eqref{eqFP}, i.e. $F(u)=0$, substituting $u_t$ by
$\frac{1}{2}\Delta u-Mu$ whenever it occurs gives an equation, of which
left hand side is a polynomial with
$u,\,u_x,\,u_y,\,u_{xx},\,u_{yy},\,u_{xy}$ and right hand side is
$0$.  Taking the coefficients of the various monomials in the first
and second order partial derivatives of $u$ in the polynomial be
$0$, we find the determining equations for the symmetry group of the
Fokker-Planck equation.

By solving them, an invariance of equation \eqref{eqFP} under
transformation \eqref{LieTrans} leads to the expressions for the
functions $T,X,Y,U$ of the form (throughout this paper we use
symbolic package MAPLE to perform all calculations)
\begin{equation}\label{vector2}\begin{cases}
   T=f_1,\\
   X=\dfrac{1}{2}\bigg(\dfrac{\partial f_1}{\partial t}\bigg)x+ky+f_2,\\
   Y=\dfrac{1}{2}\bigg(\dfrac{\partial f_1}{\partial t}\bigg)y-kx+f_3,\\
   U=-\bigg[\dfrac{1}{4}\bigg(\dfrac{\partial^2 f_1}{\partial t^2}\bigg)(x^2+y^2)
   +\bigg(\dfrac{\partial f_2}{\partial t}\bigg)x+
   \bigg(\dfrac{\partial f_3}{\partial t}\bigg)y+f_4\bigg]u+g,
\end{cases}\end{equation}
and the compatibility condition
\begin{equation}\label{condition1}
   T_tM+XM_x+YM_y+U_uM=0.
\end{equation}
where $k$ is arbitrary constant, and $f_i,\,\,i=1,\ldots,4$ are
arbitrary functions of $t$, which satisfy the condition
\eqref{condition1}, and $g$ is the solution of the original equation
\eqref{eqFP}. Similar mathematical results were given by Finkel
(1999) and Laurence and Wang (2005) by using a prolongation method.
Nowadays computer algebra packages are widely used in determination of symmetries of
differential equations. However, in this case, $T,X,Y,U$ contain the arbitrary functions of $t$
and the arbitrary function of $x,y,t$, which current software packages for symmetry analysis,
such as DESOLV, DESOLVII and SADE, cannot handle.
Therefore, we will manually use Lie symmetry analysis to deal with the Fokker-Planck equation.

The presence of these arbitrary functions leads to an
infinite-dimensional Lie algebra of symmetries. A general element of
this algebra is written as
\begin{equation}\label{algebra}
   \underline{\emph{v}}=\underline{\emph{v}}_1k+\underline{\emph{v}}_2(f_1)
   +\underline{\emph{v}}_3(f_2)+
   \underline{\emph{v}}_4(f_3)+\underline{\emph{v}}_5(f_4)+\underline{\emph{v}}_6(g).
\end{equation}

Let $\varphi_i$ be arbitrary functions of $t$, $\psi$ and $\phi$ be
arbitrary functions of $x,\,y,\,t$, then
\begin{eqnarray}
   \underline{\emph{v}}_1 & = y\frac{\partial}{\partial x}-x\frac{\partial}{\partial y},
   \nonumber\\
   \underline{\emph{v}}_2(\varphi_i) & = \varphi\frac{\partial}{\partial t}
   +\frac{1}{2}\dot{\varphi_i}x\frac{\partial}{\partial x}
   +\frac{1}{2}\dot{\varphi_i}y\frac{\partial}{\partial y}
   +\frac{1}{4}\ddot{\varphi_i}(x^2+y^2)u\frac{\partial}{\partial u},\nonumber\\
   \underline{\emph{v}}_3(\varphi_i) & = \varphi_i \frac{\partial}{\partial x}
   +\dot{\varphi_i}xu\frac{\partial}{\partial u},\nonumber\\
   \underline{\emph{v}}_4(\varphi_i) & = \varphi_i \frac{\partial}{\partial y}
   +\dot{\varphi_i}yu\frac{\partial}{\partial u},\nonumber\\
   \underline{\emph{v}}_5(\varphi_i) & = \varphi_i u\frac{\partial}{\partial u},\nonumber\\
   \underline{\emph{v}}_6(\psi) & = \psi\frac{\partial}{\partial u}.\label{symmetries}
\end{eqnarray}
The commutation relations between all these vector fields are given
by Table 1.

Table 1:\quad The commutation relations between vector fields.
\begin{itemize}
\item[]\end{itemize}
\fl\begin{tabular}{@{}l|lllllll} 
 & $\underline{\emph{v}}_1$
 & $\underline{\emph{v}}_2(\varphi_j)$
 & $\underline{\emph{v}}_3(\varphi_j)$
 & $\underline{\emph{v}}_4(\varphi_j)$
 & $\underline{\emph{v}}_5(\varphi_j)$
 & $\underline{\emph{v}}_6(\phi)$\\
\hline
 $\underline{\emph{v}}_1$            & $0$
                                     & $0$
                                     & $\underline{\emph{v}}_4(\varphi_j)$
                                     & $-\underline{\emph{v}}_3(\varphi_j)$
                                     & $0$
                                     & $\underline{\emph{v}}_6(y\phi_x-x\phi_y)$\\
 $\underline{\emph{v}}_2(\varphi_i)$ &
                                     & $\underline{\emph{v}}_2(\varphi_i\dot{\varphi_j}-\dot{\varphi_i}\varphi_j)$
                                     & $\underline{\emph{v}}_3(\varphi_i\dot{\varphi_j}-\frac{1}{2}\dot{\varphi_i}\varphi_j)$
                                     & $\underline{\emph{v}}_4(\varphi_i\dot{\varphi_j}-\frac{1}{2}\dot{\varphi_i}\varphi_j)$
                                     & $\underline{\emph{v}}_5(\varphi_i\dot{\varphi_j})$
                                     & $\underline{\emph{v}}_6(\frac{1}{2}\dot{\varphi_i}(x\phi_x+y\phi_y)+\varphi_i\phi_t)$\\
 $\underline{\emph{v}}_3(\varphi_i)$ &
                                     &
                                     & $\underline{\emph{v}}_5(\varphi_i\dot{\varphi_j}-\dot{\varphi_i}\varphi_j)$
                                     & $0$
                                     & $0$
                                     & $\underline{\emph{v}}_6(\dot{\varphi_i}x\phi-\varphi_i\phi_x)$\\
 $\underline{\emph{v}}_4(\varphi_i)$ &
                                     &
                                     &
                                     & $\underline{\emph{v}}_5(\varphi_i\dot{\varphi_j}-\dot{\varphi_i}\varphi_j)$
                                     & $0$
                                     & $\underline{\emph{v}}_6(\dot{\varphi_i}y\phi-\varphi_i\phi_y)$\\
 $\underline{\emph{v}}_5(\varphi_i)$ &
                                     &
                                     &
                                     &
                                     & $0$
                                     & $\underline{\emph{v}}_6(\varphi_i\phi)$\\
 $\underline{\emph{v}}_6(\psi)$      &
                                     &
                                     &
                                     &
                                     &
                                     & $0$\\
\end{tabular}
\begin{itemize}
\item[] The entry in row $i$ and column $j$ representing $\big[\,\underline{\emph{v}}_i\,,\,\underline{\emph{v}}_j\,\big]$.
\end{itemize}

From Table 1, we see that
${\underline{\emph{v}}_2(\varphi),\underline{\emph{v}}_3(\varphi),
\underline{\emph{v}}_4(\varphi),\underline{\emph{v}}_5(\varphi)}$
constitute a subalgebra. And there exist some types of interesting
subalgebras, For instance, Virasoro algebra and $\omega_\infty$-type
algebra.

Furthermore, we find that the transform $[\underline{\emph{v}}_1,
\underline{\emph{v}}_6(\psi)] =
\underline{\emph{v}}_6(y\psi_x-x\psi_y)$ is invariant, if $M(x,y)$
satisfies the type $C \cdot (x^2+y^2)$, where $C$ is an arbitrary constant.
In other words, if $g$ is a solution of the Fokker-Planck equation like this,
then $yg_x-xg_y$ is another solution of the same equation.

Moreover, we get a series of transformations of the solution. New
solutions can be generated through them with a known solution. The
one-parameter groups generated by $\underline{\emph{v}}_i$ and the
transformations are included in the Appendix A for the readers with
an interest in the details of applying the theory.

\section{Similarity reductions}

After determining the infinite-dimensional algebra of symmetries,
the similarity variables can be found by solving the characteristic
equations
\begin{equation}\label{reduct1}
   \frac{dt}{T}=\frac{dx}{X}=\frac{dy}{Y}=\frac{du}{U}.
\end{equation}

By solving the ordinary differential equations \eqref{reduct1}, we
can obtain integration constants  $\xi,\,\eta,\,P$. Substituting
$\xi,\,\eta,\,P$ for $x\,,y\,,t\,,u$ in original equation
\eqref{eqFP}, we can reduce the equation from $2+1-$dimensional to
$2-$dimensional finally. This process is called similarity
reductions.

Since there are many arbitrary functions in $T, \,X, \,Y, \,U$, it
is hard to solve the equations \eqref{reduct1} in the general case.
Likewise, it is also hard to solve them by substituting generators
\eqref{symmetries}. Finkel (1999) completely classified the
symmetries of the Fokker-Planck equation based on the compatibility
condition \eqref{condition1}. For simplicity, he has dropped out the
two trivial infinitesimal symmetries $\partial_t$ and $u\partial_u$
in his classification result. It means that constant terms are omitted
in the forms for $f_1$ and $f_4$.

In this subsection, we will list some cases in details for reductions as
the classification done by Finkel, which are helpful for the following subsections.
Other cases of the Fokker-Planck equation for reductions are included in Appendix B.

\begin{itemize}
\item {\bf Case 1.1a}
\end{itemize}
\begin{equation}\label{case1.1a}
\begin{cases}
   M = \dfrac{C_0}{x^2}+by+c_0,\quad C_0\neq0,\\
   f_1 = \delta_2t^2+\delta_1t,\quad k=0,\quad f_2=0,\quad f_3 = \dfrac{b\delta_2t^3}{2}
   +\dfrac{3b\delta_1t^2}{4}+\beta_1t+\beta_0,\\
   f_4 = \dfrac{b^2\delta_2t^4}{8}+\dfrac{b^2\delta_1t^3}{4}+
         \big(\dfrac{b\beta_1}{2}+c_0\delta_2\big)t^2+(\delta_2+c_0\delta_1+b\beta_0)t.
\end{cases}
\end{equation}

We have the similarity variables $\xi, \,\eta, \,P$,
\begin{equation}
\begin{cases}
   \xi = & \dfrac{x}{\sqrt{\delta_2t^2+\delta_1t}}, \\
   \eta = & \dfrac{2y\delta_1^2-b\delta_1^2t^2+4(2\beta_0\delta_2-\beta_1\delta_1)t
            +4\beta_0\delta_1}{2\delta_1^2\sqrt{\delta_2t^2+\delta_1t}}, \\
   P = & u \cdot \exp\Big\{\dfrac{1}{\delta_1^4}\{\eta\delta_1^2\sqrt{\delta_2t^2+\delta_1t}
         (\delta_1^2bt+2\delta_1\beta_1-4\delta_2\beta_0)+
         \delta_1 \ln(\delta_2t+\delta_1)(\delta_1^3+2\delta_1\beta_0\beta_1-2\delta_2\beta_0^2) \\
       & +\beta_0\delta_1 \ln(t)(2\beta_0\delta_2-2\beta_1\delta_1)+
         \dfrac{1}{3}b^2\delta_1^4t^3+2b\delta_1^2(\delta_1\beta_1-2\delta_2\beta_0)t^2 \\
       & +[8\beta_0^2\delta_2^2-8\beta_0\beta_1\delta_1\delta_2-2b\beta_0\delta_1^3
         +\delta_1^4c_0+2\beta_1^2\delta_1^2+\dfrac{1}{2}\delta_1^4\delta_2(\xi^2+\eta^2)]t\}\Big\}.
\end{cases}\label{xieta1.1a}
\end{equation}

and the reduced PDE becomes
\begin{equation}\label{eq1.1a}
   \delta_1^2\xi^2(P_{\xi\xi}+P_{\eta\eta})+\delta_1^3\xi^3P_{\xi}+\delta_1^3\xi^2\eta P_{\eta}
   +(4\delta_2\beta_0^2\xi^2-4\delta_1\beta_0\beta_1\xi^2-2C_0\delta_1^2)P=0.
\end{equation}

We get the solution by the method of separation of variables
\begin{eqnarray}
   P = {\it F_{1}}(\xi){\it F_{2}}(\eta),
\end{eqnarray}
where ${\it F_1}(\xi)$ and ${\it F_2}(\eta)$ is
\begin{equation*}\begin{cases}
   {\it F_1}(\xi) = & \frac{e^{-\frac{\delta_1 \xi^2}{4}}}{\sqrt{\xi}}
                      \bigg[C_1 {\rm WhittakerM}\Big(\frac{c_1}{2\delta_1}-\frac{1}{4},\frac{\sqrt{8C_0+1}}{4},\frac{\delta_1 \xi^2}{2}\Big)
                      +C_2 {\rm WhittakerW}\Big(\frac{c_1}{2\delta_1}-\frac{1}{4},\frac{\sqrt{8C_0+1}}{4},\frac{\delta_1 \xi^2}{2}\Big)\bigg],\\
   {\it F_2}(\eta) = & \frac{e^{-\frac{\delta_1 \eta^2}{4}}}{\sqrt{\eta}}
                       \bigg[C_3 {\rm WhittakerM}\Big(\frac{c_1}{2\delta_1}-\frac{2\beta_0\beta_1}{\delta_1^2}+\frac{2\delta_2\beta_0^2}{\delta_1^3}
                       -\frac{1}{4}, \frac{1}{4},\frac{\delta_1 \eta^2}{2}\Big) \\
                     & +C_4 {\rm WhittakerW}\Big(\frac{c_1}{2\delta_1}-\frac{2\beta_0\beta_1}{\delta_1^2}+\frac{2\delta_2\beta_0^2}{\delta_1^3}-\frac{1}{4},
   \frac{1}{4},\frac{\delta_1 \eta^2}{2}\Big)\bigg],
\end{cases}\end{equation*}
where $c_1,\,C_1,\,C_2,\,C_3,\,C_4$ are arbitrary constants.

\begin{itemize}
\item {\bf Case 1.2b}
\end{itemize}
\begin{equation}\label{case1.2b}
\begin{cases}
   M = \dfrac{C(\theta)}{r^2}+cr^2+c_0,\quad \,c\neq0,\\
   f_1 = \delta_1e^{2\sqrt{2c}t}+\delta_2e^{-2\sqrt{2c}t},\quad k=0,\quad f_2=f_3=0,\\
   f_4 = \big(\sqrt{2c}+c_0\big)\delta_1e^{2\sqrt{2c}t}-\big(\sqrt{2c}-c_0\big)\delta_2e^{-2\sqrt{2c}t}.
\end{cases}
\end{equation}
where $C(\theta)\neq (c_1\cos\theta+c_2\sin\theta)^{-2}$, $C^{'}(\theta)\neq 0$, and $r=\sqrt{x^2+y^2}$.

We have the similarity variables $\xi, \,\eta, \,P$,
\begin{equation}\label{xieta1.2b}
\begin{cases}
   \xi = & \dfrac{xe^{\sqrt{2c}t}}{\sqrt{\delta_1e^{4\sqrt{2c}t}+\delta_2}}, \\
   \eta = & \dfrac{ye^{\sqrt{2c}t}}{\sqrt{\delta_1e^{4\sqrt{2c}t}+\delta_2}}, \\
   P = & e^{-c_0t}{\sqrt{\delta_1e^{4\sqrt{2c}t}+\delta_2}} u \cdot \exp \Big\{\sqrt{\dfrac{c}{2}}
         \big[(\delta_1e^{2\sqrt{2c}t}-\delta_2e^{-2\sqrt{2c}t})(\xi^2+\eta^2)-2t\big]\Big\},
\end{cases}
\end{equation}

and the reduced PDE becomes
\begin{equation}\label{eq1.2b}
  (\xi^2+\eta^2)(P_{\xi\xi}+P_{\eta\eta})+2[4c\delta_1\delta_2(\xi^2+\eta^2)^2-C(\theta)]P=0.
\end{equation}

With the transformation $\xi=\varrho \cos\theta,\,\eta=\varrho \sin\theta$, \eqref{eq1.2b} becomes
\begin{equation}\label{eq21.2b}
   \varrho^2P_{\varrho\varrho}+\varrho P_{\varrho}+P_{\theta\theta}+2[4c\delta_1\delta_2\varrho^4-C(\theta)]P=0.
\end{equation}

We can get the solution by the method of separation of variables
\begin{eqnarray}
   P = {\it F_{1}}(\varrho){\it F_{2}}(\theta),
\end{eqnarray}
where ${\it F_1}(\varrho),\,{\it F_2}(\theta)$ is the solution of
\begin{equation}\label{ODE1.2b}\begin{cases}
   \dfrac{d^2{\it F_{1}}(\varrho)}{d\varrho^2}+\dfrac{1}{\varrho}\dfrac{d{\it F_{1}}(\varrho)}{d\varrho}
   +\big(8c\delta_{1}\delta_{2}\varrho^{2}-\dfrac{c_{1}}{\varrho^{2}}\big){\it F_{1}}(\varrho)=0,\\
   \dfrac{d^2{\it F_{2}}(\theta)}{d\theta^2}+(c_{1}-2C(\theta)){\it
   F_{2}}(\theta)=0,
\end{cases}\end{equation}
where $c_{1}$ is an arbitrary constant. Given $C(\theta)$, the ODE systems \eqref{ODE1.2b} can be solved directly.

\begin{itemize}
\item {\bf Case 1.4b}
\end{itemize}
\begin{equation}\label{case1.4b}\begin{cases}
   M = \dfrac{C_0}{r^2}+cr^2+ax+by+c_0,\\
   f_1 = \delta_1e^{2\sqrt{2c}t}+\delta_2e^{-2\sqrt{2c}t},\quad k=0,\quad f_2=f_3=0,\\
   f_4 = \big(\sqrt{2c}+c_0\big)\delta_1e^{2\sqrt{2c}t}-\big(\sqrt{2c}-c_0\big)
   \delta_2e^{-2\sqrt{2c}t},
\end{cases}\end{equation}
where $r=\sqrt{x^2+y^2}$. If $\delta_1\neq0,\,\delta_2\neq0$,
then $a=b=0$ should be held to make $M$ satisfy the compatibility condition \eqref{condition1}.
Obviously, this is the simplification of \textbf{Case 1.2b}.
We have the similarity variables $\xi, \,\eta, \,P$ same as \eqref{xieta1.2b}, then the reduced PDE becomes
\begin{equation}\label{eq1.4b}
   (\xi^2+\eta^2)(P_{\xi\xi}+P_{\eta\eta})+2[4c\delta_1\delta_2(\xi^2+\eta^2)^2-C_0]P=0.
\end{equation}

With the transformation $\xi=\varrho \cos\theta,\,\eta=\varrho \sin\theta$,
we can get the solution by the method of separation of variables
\begin{eqnarray}
   P = {\it F_{1}}(\varrho){\it F_{2}}(\theta),
\end{eqnarray}
where ${\it F_1}(\varrho)$ and ${\it F_2}(\theta)$ is
\begin{equation*}\begin{cases}
   {\it F_1}(\varrho) = & C_1 J\Big(\dfrac{\sqrt{c_1}}{2},\sqrt{2\delta_1\delta_2c}\varrho^2\Big)
                          +C_2 Y\Big(\dfrac{\sqrt{c_1}}{2},\sqrt{2\delta_1\delta_2c}\varrho^2\Big)\\
   {\it F_2}(\theta) = & C_3 \sin(\theta\sqrt{c_1-2C_0})+C_4 \cos(\theta\sqrt{c_1-2C_0}),
\end{cases}\end{equation*}
where $c_1,\,C_1,\,C_2,\,C_3,\,C_4$ are arbitrary constants, and
$J(\nu, z)$ and $Y(\nu, z)$ are the Bessel functions of the first
and second kinds, respectively.

\section{Case studies}

We now study a few cases, most of which are not well known in the
financial literature. Our purpose here is to demonstrate the
procedure of producing analytical option pricing formulas with the
method of similarity reduction.

\subsection{$2-$dimensional: Double CEV Model}

In the traditional CEV Model (Cox 1975, Cox and Ross 1976, Schroder
1989), $\sigma(S)=\sigma S^{\alpha}$. Base on their work, we try to
build a Double CEV Model, which has two assets. Assuming
\begin{equation}
   \sigma_i(S_i)=\sigma_i S_i^{\alpha_i},
\end{equation}\label{dcev}
where $\sigma_i>0$, $\alpha_i >0$. From \eqref{trans1}, we have
\begin{equation}\label{dcevtrans1}
\begin{cases}
   \displaystyle x = \sqrt{\frac{2}{1+\rho}}\big(\dfrac{S_1^{1-\alpha_1}}{\sigma_1(1-\alpha_1)}
                     +\dfrac{S_2^{1-\alpha_2}}{\sigma_2(1-\alpha_2)}\big),\\
   \displaystyle y = \sqrt{\frac{2}{1-\rho}}\big(\dfrac{S_2^{1-\alpha_2}}{\sigma_2(1-\alpha_2)}
                     -\dfrac{S_1^{1-\alpha_1}}{\sigma_1(1-\alpha_1)}\big)),
\end{cases}
\end{equation}
and the following transformation \eqref{trans2}, where
\begin{equation*}
\begin{cases}
  Q_1 = & \dfrac{r(1-\alpha_1)}{2}\Big(x-\sqrt{\frac{1-\rho}{1+\rho}}y\Big)
          +\dfrac{r(1-\alpha_2)}{2}\Big(x+\sqrt{\frac{1-\rho}{1+\rho}}y\Big) \\
        & -\dfrac{\alpha_1}{1-\alpha_1}\dfrac{1}{\Big(\frac{1+\rho}{2}x-\frac{\sqrt{1-\rho^2}}{2}y\Big)}
          -\dfrac{\alpha_2}{1-\alpha_2}\dfrac{1}{\Big(\frac{1+\rho}{2}x+\frac{\sqrt{1-\rho^2}}{2}y\Big)}, \\
  Q_2 = & \dfrac{r(1-\alpha_2)}{2}\Big(\sqrt{\frac{1+\rho}{1-\rho}}x+y\Big)
          -\dfrac{r(1-\alpha_1)}{2}\Big(\sqrt{\frac{1+\rho}{1-\rho}}x-y\Big) \\
        & -\dfrac{\alpha_1}{1-\alpha_1}\dfrac{1}{\Big(\frac{\sqrt{1-\rho^2}}{2}x-\frac{1-\rho}{2}y\Big)}
          -\dfrac{\alpha_2}{1-\alpha_2}\dfrac{1}{\Big(\frac{\sqrt{1-\rho^2}}{2}x+\frac{1-\rho}{2}y\Big)},
\end{cases}
\end{equation*}
and
\begin{equation*}
   \omega(x,y)=2\Big(\frac{\alpha_1ln(x-y)}{1-\alpha_1}+\frac{\alpha_2ln(x+y)}{1-\alpha_2}\Big)+
               \frac{r}{4}\big[\alpha_1(x-y)^{2}+\alpha_2(x+y)^{2}-2(x^{2}+y^{2})\big]+C_0,
\end{equation*}
where $C_0$ is an arbitrary constant.

Equation \eqref{eq0} becomes
\begin{equation}
   u_t(x,y,t)-\frac{1}{2}\Delta u(x,y,t)+M(x,y,t)u=0,
\end{equation}
where
\begin{equation}
   M(x,y,t) = \frac{48(x^2+y^2)}{(x^2-y^2)^2}+r^2(x^2+y^2)-18r.
\end{equation}
For brevity we have taken $\rho=0$$,\,\alpha_1=\alpha_2=2$.\\
(If $\rho\neq0$, the compatibility condition \eqref{condition1} is
also satisfied with the following $f_i,\,C$ and $g$. Moreover, the
similarity variables and the solution of the reduced PDE can be
obtained. Here, taking $\rho=0$ is just for brevity.)

Obviously, the function $M$ belongs to the {\bf Case 1.2b}.
Therefore, we take
\begin{equation*}\begin{cases}
   f_1 = \delta_1e^{2\sqrt{2}rt}+\delta_2e^{-2\sqrt{2}rt},\quad k=0,\quad f_2=f_3=0,\\
   f_4 = \big(\sqrt{2}r-18r\big)\delta_1e^{2\sqrt{2}rt}-\big(\sqrt{2}r+18r\big)\delta_2e^{-2\sqrt{2}rt}.
\end{cases}\end{equation*}

We have the similarity variables $\xi, \,\eta, \,P$,
\begin{equation}
\begin{cases}
   \xi = & \dfrac{xe^{\sqrt{2}rt}}{\sqrt{\delta_1e^{4\sqrt{2}rt}+\delta_2}}, \\
   \eta = & \dfrac{ye^{\sqrt{2}rt}}{\sqrt{\delta_1e^{4\sqrt{2}rt}+\delta_2}}, \\
   P = & e^{-18rt}{\sqrt{\delta_1e^{4\sqrt{2}rt}+\delta_2}} u \cdot \exp \Big\{\dfrac{1}{\sqrt{2}}r
         \big[(\delta_1e^{2\sqrt{2}rt}-\delta_2e^{-2\sqrt{2}rt})(\xi^2+\eta^2)-2t\big]\Big\},
\end{cases}\label{dcevtrans}
\end{equation}

and the reduced PDE becomes
\begin{equation}\label{dceveq1}
   (P_{\xi\xi}+P_{\eta\eta})+8r^2\delta_1\delta_2(\xi^2+\eta^2)P-\frac{96(\xi^2+\eta^2)}{(\xi^2-\eta^2)^2}P=0.
\end{equation}

With the transformation $\xi=\varrho \cos\theta,\,\eta=\varrho \sin\theta$, \eqref{dceveq1} becomes
\begin{equation}\label{dceveq2}
   \varrho^2P_{\varrho\varrho}+\varrho P_{\varrho}+P_{\theta\theta}+8r^2\delta_1\delta_2\varrho^4P-\frac{96P}{\cos^22\theta}=0.
\end{equation}

The solution can be written as
\begin{eqnarray}
   P = {\it F_{1}}(\varrho){\it F_{2}}(\theta),\label{dcevsoution}
\end{eqnarray}
where ${\it F_1}(\varrho)$ and ${\it F_2}(\theta)$ are
\begin{equation*}\begin{cases}
   {\it F_1}(\varrho) = & C_1 J\Big(\dfrac{\sqrt{c_1}}{2},\sqrt{2\delta_1\delta_2}r\varrho^2\Big)
                          +C_2 Y\Big(\dfrac{\sqrt{c_1}}{2},\sqrt{2\delta_1\delta_2}r\varrho^2\Big)\\
   {\it F_2}(\theta) = & \dfrac{(2 \cos(4\theta)-2)^{\frac{3}{4}}}{\sqrt{ \sin(4\theta)}}\bigg\{
                         C_3(\frac{ \cos(4\theta)+1}{2})^{\frac{1}{2}+\frac{\sqrt{97}}{4}} {\rm Hypergeom} \Big(
                         \big[\frac{3+\sqrt{97}+\sqrt{c_1}}{4},\frac{3+\sqrt{97}-\sqrt{c_1}}{4}\big],\\
                       & \big[1+\frac{\sqrt{97}}{2}\big],\frac{\cos(4\theta)+1}{2}\Big)
                         +C_4(\frac{\cos(4\theta)+1}{2})^{\frac{1}{2}-\frac{\sqrt{97}}{4}} {\rm Hypergeom} \Big(
                         \big[\frac{3-\sqrt{97}+\sqrt{c_1}}{4},\frac{3-\sqrt{97}-\sqrt{c_1}}{4}\big],\\
                       & \big[1-\frac{\sqrt{97}}{2}\big],\frac{\cos(4\theta)+1}{2}\Big)\bigg\},
\end{cases}\end{equation*}
where $c_1,\,C_1,\,C_2,\,C_3,\,C_4$ are arbitrary constants, and
Hypergeom is generalized hypergeometric function.

We can get the original solution $c$ of generalized Black-Scholes equation \eqref{eq0} through substituting $P$ with the
transformation \eqref{dcevtrans}, \eqref{trans2} and \eqref{trans1}.

For $1-$dimensional generalized Black-Scholes equation, similarity
reduction method can be used to reduce the PDE to an ODE which is
easier to solve. Except the time dependent cases, we can also use
this method to reduce all equations Carr, Laurence and Wang (2006)
transformed, which are associated with the $1-$dimensional
simplification of {\bf Case 1.4b}.

\subsection{$1-$dimensional: CEV Model}

Assuming
\begin{equation}
   \sigma(S)=\sigma S^{\alpha},
\end{equation}\label{cev}
from the transformation, we know that the corresponding $M(x)$
\begin{equation}
   M=\alpha\sigma^2\big(\frac{\alpha\sigma^2}{(\alpha-1)^2}-\frac{1}{2(\alpha-1)}\big)
   \frac{1}{x^2}
     +\frac{r^2(\alpha-1)^2}{\sigma^4}x^2-2r\alpha-\frac{r(\alpha-1)}{2\sigma^2},
\end{equation}\label{cevM}
is the one dimensional case of {\bf Case 1.4b}. Therefore the solution can be written as
\begin{equation}
   P(\xi) = c_1 \xi^{\frac{1}{4}-\frac{c_0}{\sqrt{2c}}}.
\end{equation}\label{Scev}

We can get the original solution $c$ by substituting $P$ with the transformation \eqref{xieta1.2b}, \eqref{trans2} and
\eqref{trans1} ($1-$dimensional form).

\subsection{$1-$dimensional: Exponentially Decreasing Volatility}

Assuming
\begin{equation}\label{edv}
   \sigma(S)=e^{-S},
\end{equation}
from the transformation, we know that the corresponding $M(x)$
\begin{equation}\label{edvM}
   M=\frac{1}{2x^2},
\end{equation}
where for brevity we let $r=0$ \footnote{Similar transformation can
be found in Li and Zhang (2004), and Zhang and Li (2012).}, is the
one dimensional case of {\bf Case 1.1a}. Therefore the solution can
be written as
\begin{equation}
   P(\xi) = \frac{e^{-\frac{\delta_1 \xi^2}{4}}}{\sqrt{\xi}}
   \big[c_1 {\rm WhittakerM}(\frac{\delta_2t}{2\delta_1}-\frac{1}{4},\frac{\sqrt{5}}{4},
   \frac{\delta_1 \xi^2}{2})
   +c_2 {\rm WhittakerW}(\frac{\delta_2t}{2\delta_1}-\frac{1}{4},\frac{\sqrt{5}}{4},
   \frac{\delta_1 \xi^2}{2})\big].
\end{equation}\label{Sedv}
where WhittakerM and WhittakerW are the Whittaker function $M$ and
$W$, respectively.

We can get the original solution $c$ through substituting $P$ with
the transformation \eqref{xieta1.1a}, \eqref{trans2} and
\eqref{trans1} ($1-$dimensional form).

With a proper re-scaling transformation, the exponential decreasing
function volatility function can be converted to
$$ \sigma(S_t) = \sigma_0 S_0 e^{\alpha \left(1-\frac{S_t}{S_0} \right)},$$
where $S_0$ stands for the initial stock price, then
\bea \tilde \sigma(S_t) &=& \sigma_0 \frac{S_0}{S_t} e^{\alpha
\left(1-\frac{S_t}{S_0} \right)} \nn \\
&=& \sigma_0 \left \{ 1 -(1+\alpha) \left(\frac{S_t}{S_0} - 1
\right) + \left( 1 + \alpha + \frac 1 2 \alpha^2 \right)
\left(\frac{S_t}{S_0} - 1 \right)^2 + O \left[ \left(\frac{S_t}{S_0}
- 1 \right) \right]^3 \right \}, \nn \nda
where $O(\epsilon)$ is the order of $\epsilon$. The function is
negatively skewed for positive $\alpha$, can be used to produce the
phenomenon of the implied volatility smirk observed by Zhang and
Xiang (2008), see also, Zhang and Li (2012).

\section{Conclusion}

With some transformation, we convert the problem of option pricing
under state-dependent volatility into an initial value problem of
the Fokker-Planck equation with a certain potential. By using the
Lie symmetry analysis and similarity reduction method, we are able
to write the solution analytically.

The study on a few cases demonstrates that our new method can be
used to produce analytical option pricing formulas for certain
volatility functions. A few exact solutions of the corresponding
cases provided in this paper can be regarded as contributions to the
option pricing literature.

The comparison with Finkel (1999), and Laurence and Wang (2005) is
as follows. In terms of the method, Finkel (1999) studied $2+1-$
dimensional Fokker-Planck equation in general by using the
prolongation of vector-field, but he did not discuss the
applications in finance. Laurence and Wang (2005) used the same
method as Finkel's and applied Finkel's results in finance. Our
method presented in Section 3 is more succinct. In terms of the
results, Finkel (1999) provided the vector fields of group
invariants in the symmetry reduction and group invariant solutions
in the particular case of 1.1a. 
On the top of Finkel (1999), Laurence and Wang
(2005) provided the group invariant solutions via subgroups
generated by particular subalgebras in cases of 1.1ab, 1.2ab, 1.4ab,
1.5ab, 1.7ab. 
We perform similarity reduction, and provide the group invariant
solutions in the cases of 1.3, 1.6 and 1.8ab. In terms of finance
application, Laurence and Wang (2005) only studied the case of the
generalized Black-Scholes equation on a single asset. We point out
that the problem of independent double-CEV can be reduced to case
1.2b. Even for the case of a single asset, the examples in our
Section 5.2, 5.3, were not studied in Laurence and Wang (2005).

In finance, it is an open problem to find a closed form solution for
the option on two correlated CEV assets. The case of independent
double-CEV has been studied in Section 5.1 in this paper. In order
to study the case of correlated double CEV, we need to use the
general case of Fokker-Planck equation. In principle, we can find
generators of invariant groups by using Lie symmetry approach. With
Finkel's (1999) classification, we can then obtain reduced equation
by using similarity reduction method. In quantitative finance, we
are interested in a solution of the generalized Black-Scholes
equation with a particular final condition, i.e., payoff function.
Constructing a solution of relevance in quantitative finance by
using some particular solutions seems not straightforward. This
problem is left for further research.

It is also an interesting topic to explore the application of
current approach to the pricing of path-dependent derivatives.

\section*{Acknowledgement.}
Wenqing Bao has been supported by the Soft Science Research Program
of Shanghai Science and Technology Development Fund (No. 201006007).
Jin E. Zhang has been supported by an establishment grant from
University of Otago.

\pagebreak

\appendix
\section{Transforms of the solution}

Given a vector field $\underline{\emph{v}}$, the corresponding
one-parameter group of infinitesimal transformation
$G:(x,\,y,\,t,\,u) \rightarrow
(\bar{x},\,\bar{y},\,\bar{t},\,\bar{u})$ can be obtained by solving
the ODE
\begin{equation*}
\begin{cases}
  \dfrac{d}{d\epsilon}(\bar{x},\,\bar{y},\,\bar{t},\,\bar{u}) = (X,\,Y,\,T,\,U)
  (\bar{x},\,\bar{y},\,\bar{t},\,\bar{u}),\\
  (\bar{x},\,\bar{y},\,\bar{t},\,\bar{u})|_{\epsilon=0} = (x,\,y,\,t,\,u).
\end{cases}
\end{equation*}

They are
\begin{eqnarray}
  &G_1:(x,\,y,\,t,\,u) & \rightarrow (x\cos(\epsilon)+y\sin(\epsilon),\,
  -x\sin(\epsilon)+y\cos(\epsilon),\,t,\,u), \nonumber\\
  &G_2:(x,\,y,\,t,\,u) & \rightarrow (\bar{x},\,\bar{y},\,\bar{t},\,\bar{u}), \nonumber\\
  &G_3:(x,\,y,\,t,\,u) & \rightarrow (x+f_2\epsilon,\,y,\,t,\,ue^{f_{2t}(\frac{1}{2}f_2\epsilon^2
  +x\epsilon)}), \nonumber\\
  &G_4:(x,\,y,\,t,\,u) & \rightarrow (x,\,y+f_3\epsilon,\,t,\,ue^{f_{3t}(\frac{1}{2}f_3\epsilon^2
  +y\epsilon)}), \nonumber\\
  &G_5:(x,\,y,\,t,\,u) & \rightarrow (x,\,y,\,t,\,ue^{f_4\epsilon}), \nonumber\\
  &G_6:(x,\,y,\,t,\,u) & \rightarrow (x,\,y,\,t,\,u+g\epsilon), \nonumber
\end{eqnarray}
where $\epsilon$ is an arbitrary constant, and
$f_i,\,\,i=1,\ldots,4$ are arbitrary functions of $t$, which satisfy
the compatibility condition \eqref{condition1}, and $g$ is the
solution of the original equation \eqref{eqFP}. Solving $G_2$ is
feasible only when given the definite form of $f_1$. We have tried
to solve it with two forms (polynomial function and exponential
function) \footnote{They are associated with $f_1$ in Section 4 and
Appendix B.} :
\begin{eqnarray}
   f_1 = \delta_2 t^2+\delta_1 t,\quad f_1 = \delta_1e^{2\sqrt{2c}t}+\delta_2e^{-2\sqrt{2c}t}.
   \nonumber
\end{eqnarray}

Due to the space limitation, we only consider a special case here: $f_1 = \delta t$.
\begin{eqnarray}
  G_2:(x,\,y,\,t,\,u) \rightarrow (xe^{\frac{1}{2}e^{\delta\epsilon}(\delta t-1)},\,
                                   ye^{\frac{1}{2}e^{\delta\epsilon}(\delta t-1)},\,
                                   te^{\delta\epsilon},\,
                                   ue^{\frac{1}{4}e^{\delta t(e^{\delta t}-1)-1}\delta(x^2+y^2)}), \nonumber
\end{eqnarray}

We observe that $G_1$ is a rotation, $G_3$ and $G_4$ are
compositions of space translation and Galileo boost, $G_5$ is a
Galileo boost, $G_6$ shows that the solution of original equation
\eqref{eqFP} is linear, which is consistent with the fact that the
equation itself is linear, $G_2$ is a Galileo boost when $f_1 =
\delta t$. The entire symmetry group is obtained by combining the
six subgroups $G_i,\, i=1,\ldots,6$.

Furthermore, if $u=\phi(x,\,y,\,t)$ is the solution of Fokker-Planck equation, then so are the functions
$u^{(1)},\,u^{(2)},\ldots,u^{(6)}$,
\begin{eqnarray}
  &u^{(1)} & = \phi(x\cos(\epsilon)-y\sin(\epsilon),\,x\sin(\epsilon)+y\cos(\epsilon),\,t),
  \nonumber\\
  &u^{(2)} & = e^{\frac{1}{4}\delta(x^2+y^2)e^{\delta t(e^{-\delta\epsilon}-1)}}
              \phi(xe^{\frac{1}{2}\delta t(e^{-\delta\epsilon}-1)},\,ye^{\frac{1}{2}\delta
              t(e^{-\delta\epsilon}-1)},
              \,te^{-\delta\epsilon}), \nonumber\\
  &u^{(3)} & = e^{f_{2t}(\frac{1}{2}f_2\epsilon^2-x\epsilon)} \phi(x-f_2\epsilon,\,y,\,t), \nonumber\\
  &u^{(4)} & = e^{f_{3t}(\frac{1}{2}f_3\epsilon^2-y\epsilon)} \phi(x,\,y-f_3\epsilon,\,t), \nonumber\\
  &u^{(5)} & = e^{-f_4\epsilon} \phi(x,\,y,\,t), \nonumber\\
  &u^{(6)} & = \phi(x,\,y,\,t)-g\epsilon. \nonumber
\end{eqnarray}

By using some one-parameter groups of transformation $G_i$, a new solution can be generated. Moreover, we can use groups
$G_1,\,G_2,\ldots,G_6$ compositely by taking different constant $\epsilon_1,\,\epsilon_2,\ldots,\epsilon_6$, to obtain a series
of new solutions.

\pagebreak
\section{Similarity reductions of Fokker-Planck equation}

\begin{itemize}
\item {\bf Case 1.1b}
\end{itemize}
\begin{equation*}
\begin{cases}
   M = \dfrac{C_0}{x^2}+cr^2+by+c_0,\quad C_0\neq0,\,c\neq0,\\
   f_1 = \delta_1e^{2\sqrt{2c}t}+\delta_2e^{-2\sqrt{2c}t},\quad k=0,\quad f_2=0,\\
   f_3 = \dfrac{b\delta_1}{\sqrt{2c}}e^{2\sqrt{2c}t}-\dfrac{b\delta_2}{\sqrt{2c}}e^{-2\sqrt{2c}t}
         +\beta_1e^{\sqrt{2c}t}+\beta_2e^{-\sqrt{2c}t},\\
   f_4 = \big(\sqrt{2c}+c_0+\dfrac{b^2}{4c}\big)\delta_1e^{2\sqrt{2c}t}-\big(\sqrt{2c}-c_0
   -\dfrac{b^2}{4c}\big)\delta_2e^{-2\sqrt{2c}t}
         +\dfrac{b\beta_1}{\sqrt{2c}}e^{\sqrt{2c}t}-\dfrac{b\beta_2}{\sqrt{2c}}
         e^{-\sqrt{2c}t}.
\end{cases}
\end{equation*}
where $r=\sqrt{x^2+y^2}$.

We have the similarity variables $\xi, \,\eta, \,P$,
\begin{equation*}
\begin{cases}
   \xi = & \dfrac{xe^{\sqrt{2c}t}}{4\sqrt{\delta_1e^{4\sqrt{2c}t}+\delta_2}}, \\
   \eta = & \dfrac{xe^{\sqrt{2c}t}(4yc\delta_1\delta_2+2b\delta_1\delta_2
            -\delta_1\beta_2\sqrt{2c}e^{\sqrt{2c}t}+\delta_2\beta_1\sqrt{2c}e^{-\sqrt{2c}t})}
            {c\delta_1\delta_2\sqrt{\delta_1e^{4\sqrt{2c}t}+\delta_2}}, \\
   P = & u \cdot \exp\bigg\{\frac{1}{16c^{\frac{3}{2}}(\delta_1\delta_2)^{\frac{5}{2}}e^{\sqrt{2c}t}}\Big[
         8\eta c^{\frac{3}{2}}(\delta_1\delta_2)^{\frac{3}{2}}\sqrt{\delta_1e^{4\sqrt{2c}t}+\delta_2}
         (\beta_1\delta_2+\beta_2\delta_1e^{2\sqrt{2c}t}) \\
       & -2\sqrt{2}c\delta_1\delta_2(\beta_1^2\delta_2+\beta_2^2\delta_1)\arctan\big(\sqrt{\frac{\delta_1}{\delta_2}}e^{2\sqrt{2c}t}\big) \\
       & +4\sqrt{c}(\delta_1\delta_2)^{\frac{5}{2}}t(4c_0-4\sqrt{2}c-b^2)e^{2\sqrt{2c}t} \\
       & +\sqrt{2}c\delta_1^{\frac{5}{2}}\delta_2^{\frac{1}{2}[\beta_2^2+8c\delta_1\delta_2^2}(\xi^2+\eta^2)]e^{4\sqrt{2c}t}
         -\sqrt{2}c\delta_1^{\frac{1}{2}}\delta_2^{\frac{5}{2}[\beta_1^2+8c\delta_1^2\delta_2}(\xi^2+\eta^2)]\Big]\bigg\},
\end{cases}
\end{equation*}

and the reduced PDE becomes
\begin{equation*}
   \delta_1\delta_2\xi^2(P_{\xi\xi}+P_{\eta\eta})+[8c\delta_1^2\delta_2^2\xi^2(\xi^2+\eta^2)
   -\xi^2(\beta_1^2\delta_2+\beta_2^2\delta_1)-2C_0\delta_1\delta_2]P=0.
\end{equation*}

We get the solution by the method of separation of variables
\begin{eqnarray}
   P = {\it F_{1}}(\xi){\it F_{2}}(\eta),
\end{eqnarray}
where ${\it F_1}(\xi)$ and ${\it F_2}(\eta)$ is
\begin{equation*}\begin{cases}
   {\it F_1}(\xi) = & \frac{1}{\sqrt{\xi}}
                      \bigg[C_1 {\rm WhittakerM}\Big(i\frac{\sqrt{2}c_1}{16\sqrt{\delta_1\delta_2c}},\frac{\sqrt{8C_0+1}}{4},
                                                i2\sqrt{2\delta_1\delta_2c}\xi^2\Big) \\
                    &  +C_2 {\rm WhittakerW}\Big(i\frac{\sqrt{2}c_1}{16\sqrt{\delta_1\delta_2c}},\frac{\sqrt{8C_0+1}}{4},
                                           i2\sqrt{2\delta_1\delta_2c}\xi^2\Big)\bigg],\\
   {\it F_2}(\eta) = & \frac{1}{\sqrt{\eta}}
                       \bigg[C_3 {\rm WhittakerM}\Big(i\frac{\sqrt{2}(\beta_1^2\delta_2-\beta_2^2\delta_1-\delta_1\delta_2c_1)}{16\delta_1^{3/2}\delta_2^{3/2}\sqrt{c}},
                                                 \frac{1}{4},i2\sqrt{2\delta_1\delta_2c}\eta^2\Big) \\
                     & +C_4 {\rm WhittakerW}\Big(i\frac{\sqrt{2}(\beta_1^2\delta_2-\beta_2^2\delta_1-\delta_1\delta_2c_1)}{16\delta_1^{3/2}\delta_2^{3/2}\sqrt{c}},
                                            \frac{1}{4},i2\sqrt{2\delta_1\delta_2c}\eta^2\Big)\bigg],
\end{cases}\end{equation*}
where $c_1,\,C_1,\,C_2,\,C_3,\,C_4$ are arbitrary constants, and
WhittakerM and WhittakerW are the Whittaker function $M$ and $W$,
respectively, and $i=\sqrt{-1}$.

\begin{itemize}
\item {\bf Case 1.2a}
\end{itemize}
\begin{equation*}
\begin{cases}
   M = \dfrac{C(\theta)}{r^2}+c_0,\\
   f_1 = \delta_2t^2+\delta_1t,\quad k=0,\quad f_2=f_3=0,\\
   f_4 = c_0\delta_2t^2+(\delta_2+c_0\delta_1)t.
\end{cases}
\end{equation*}
where $C(\theta)\neq (c_1\cos\theta+c_2\sin\theta)^{-2}$, $C^{'}(\theta)\neq 0$, and $r=\sqrt{x^2+y^2}$.

We have the similarity variables $\xi, \,\eta, \,P$,
\begin{equation*}
\begin{cases}
   \xi = & \dfrac{x}{\sqrt{\delta_2t^2+\delta_1t}}, \\
   \eta = & \dfrac{y}{\sqrt{\delta_2t^2+\delta_1t}}, \\
   P = & (\delta_2t+\delta_1)u \cdot
   \exp\{-\dfrac{\delta_2}{2}(\xi^2+\eta^2)t-c_0t\},
\end{cases}
\end{equation*}

and the reduced PDE becomes
\begin{equation*}
   (\xi^2+\eta^2)(P_{\xi\xi}+P_{\eta\eta})+\delta_1(\xi^2+\eta^2)(\xi P_{\xi}+\eta P_{\eta})-2C(\theta)P=0.
\end{equation*}

With the transformation $\xi=\varrho \cos\theta,\,\eta=\varrho \sin\theta$, it becomes
\begin{equation*}
   \varrho^2P_{\varrho\varrho}+(\delta_1\varrho^3+\varrho) P_{\varrho}+P_{\theta\theta}-2C(\theta)]P=0.
\end{equation*}

We can get the solution by the method of separation of variables
\begin{eqnarray}
   P = {\it F_{1}}(\varrho){\it F_{2}}(\theta),\nonumber
\end{eqnarray}
where ${\it F_1}(\varrho),\,{\it F_2}(\theta)$ is the solution of
\begin{equation*}\begin{cases}
   \dfrac{d^2{\it F_{1}}(\varrho)}{d\varrho^2}+\dfrac{\delta_1\varrho^2+1}{\varrho}\dfrac{d{\it F_{1}}(\varrho)}{d\varrho}
   -\dfrac{c_1{\it F_{1}}(\varrho)}{\varrho^2}=0,\\
   \dfrac{d^2{\it F_{2}}(\theta)}{d\theta^2}+(c_{1}-2C(\theta)){\it F_{2}}(\theta)=0.
\end{cases}\end{equation*}
where $c_{1}$ is an arbitrary constant. Given $C(\theta)$, the above ODE systems can be solved directly.

\begin{itemize}
\item {\bf Case 1.3}
\end{itemize}
\begin{equation*}
\begin{cases}
   M = \dfrac{C(\lambda \ln r+\theta)}{r^2}+c_0,\quad C_0\neq0,\\
   f_1 = \dfrac{2k}{\lambda}t,\quad f_2=f_3=0,\quad f_4 = \dfrac{2kc_0}{\lambda}t.
\end{cases}
\end{equation*}
where $C^{'}(\theta)\neq 0 \neq \lambda$, and $r=\sqrt{x^2+y^2}$.

We have the similarity variables $\xi, \,\eta, \,P$,
\begin{equation*}
\begin{cases}
   \xi = & -\sqrt{\frac{1}{t}}\big(x\cos(\frac{1}{2}\lambda \ln t)-y\sin(\frac{1}{2}\lambda \ln t)\big), \\
   \eta = & \sqrt{\frac{1}{t}}\big(x\sin(\frac{1}{2}\lambda \ln t)+y\cos(\frac{1}{2}\lambda \ln t)\big), \\
   P = & u e^{c_0t}.
\end{cases}
\end{equation*}

and the reduced PDE becomes
\begin{equation*}
  (\xi^2+\eta^2)[P_{\xi\xi}+P_{\eta\eta}+(\xi-\lambda\eta)p_{\xi}+(\eta+\lambda\xi)p_{\eta}]
  -2C(\lambda lnr+\theta)P=0.
\end{equation*}

\begin{itemize}
\item {\bf Case 1.4a}
\end{itemize}
\begin{equation*}
\begin{cases}
   M = \dfrac{C_0}{r^2}+ax+by+c_0,\quad C_0\neq0,\\
   f_1 = \delta_2t^2+\delta_1t,\quad k=0,\quad f_2=f_3=0,\\
   f_4 = c_0\delta_2t^2+(\delta_2+c_0\delta_1)t.
\end{cases}
\end{equation*}
where $r=\sqrt{x^2+y^2}$. If $\delta_1\neq0,\,\delta_2\neq0$,
then $a=b=0$ should be held to make $M$ satisfy the compatibility condition \eqref{condition1}.

Obviously, this is the simplification of \textbf{Case 1.2a}. We have the same similarity variables $\xi, \,\eta, \,P$,
then the reduced PDE becomes
\begin{equation*}
   (\xi^2+\eta^2)(P_{\xi\xi}+P_{\eta\eta})+\delta_1(\xi^2+\eta^2)(\xi P_{\xi}+\eta P_{\eta})-2C_0P=0.
\end{equation*}

With the transformation $\xi=\varrho \cos\theta,\,\eta=\varrho \sin\theta$,
we can get the solution by the method of separation of variables
\begin{eqnarray}
   P = {\it F_{1}}(\varrho){\it F_{2}}(\theta),\nonumber
\end{eqnarray}
where ${\it F_1}(\varrho)$ and ${\it F_2}(\theta)$ is
\begin{equation*}\begin{cases}
   {\it F_1}(\varrho) = & \varrho e^{-\dfrac{\delta_1\varrho^2}{4}}\bigg[C_1 I\Big(\dfrac{\sqrt{c_1}-1}{2},\dfrac{\delta_1\varrho^2}{4}\Big)
                          +C_1 I\Big(\dfrac{\sqrt{c_1}+1}{2},\dfrac{\delta_1\varrho^2}{4}\Big)\\
                        & +C_2 K\Big(\dfrac{\sqrt{c_1}-1}{2},\dfrac{\delta_1\varrho^2}{4}\Big)
                          -C_2 K\Big(\dfrac{\sqrt{c_1}+1}{2},\dfrac{\delta_1\varrho^2}{4}\Big)\bigg],\\
   {\it F_2}(\theta) = & C_3 \sin(\theta\sqrt{c_1-2C_0})+C_4 \cos(\theta\sqrt{c_1-2C_0}),
\end{cases}\end{equation*}
where $c_1,\,C_1,\,C_2,\,C_3,\,C_4$ are arbitrary constants, and
$I(\nu, z)$ and $K(\nu, z)$ are the modified Bessel functions of the
first and second kinds respectively.

\begin{itemize}
\item {\bf Case 1.5a}
\end{itemize}
\begin{equation*}
\begin{cases}
   M = ax+by+c_0,\\
   f_1 = \delta_2t^2+\delta_1,\\
   f_2 = \dfrac{a\delta_2}{2}t^3+\dfrac{1}{4}(3a\delta_1-2bk)t^2+\alpha_1t+\alpha_0,\\
   f_3 = \dfrac{b\delta_2}{2}t^3+\dfrac{1}{4}(3b\delta_1+2ak)t^2+\beta_1t+\beta_0, \\
   f_4 = \dfrac{1}{8}(a^2+b^2)\delta_2t^4+\dfrac{1}{4}(a^2+b^2)\delta_1t^3+\big[\dfrac{1}{2}(a\alpha_1+b\beta_1)+c_0\delta_2\big]t^2
         +(\delta_2+c_0\delta_1+a\alpha_0+b\beta_0)t.
\end{cases}
\end{equation*}
Taking $k=0$ for brevity, we have the similarity variables $\xi, \,\eta, \,P$,
\begin{equation*}
\begin{cases}
   \xi = & \dfrac{(2x-a^2t)\delta_1^2+4(\alpha_0-\alpha_1t)\delta_1+8\alpha_0\delta_2t}{2\sqrt{\delta_2t^2+\delta_1}}, \\
   \eta = & \dfrac{(2y-b^2t)\delta_1^2+4(\beta_0-\beta_1t)\delta_1+8\beta_0\delta_2t}{2\sqrt{\delta_2t^2+\delta_1}}, \\
   P = & u \cdot \exp\bigg\{\dfrac{1}{\delta_1^4}\Big\{\delta_1^2\big\{\delta_1^2(a\xi+b\eta)t+2\delta_1(\alpha_1\xi+\beta_1\eta)
         -4\delta_2(\alpha_0\xi+\beta_0\eta)\big\}\sqrt{\delta_2t^2+\delta_1t} \\
       & +\delta_1\big\{\delta_1^3+2\delta_1(\alpha_0\alpha_1+\beta_0\beta_1)-2\delta_2(\alpha_0^2+\beta_0^2)\big\}ln(\delta_2t+\delta_1)
         -2\delta_1\big[\delta_1(\alpha_0\alpha_1+\beta_0\beta_1) \\
       & -\delta_2(\alpha_0^2+\beta_0^2)\big]lnt+\frac{t}{6}\big\{2\delta_1^4t^2(a^2+b^2)+3\delta_1\delta_2(\xi^2+\eta^2)+6\delta_1^4c_0 \\
       & +12\delta_1^2[\delta_1t(a\alpha_1+b\beta_1)-\delta_1(a\alpha_0+b\beta_0)+\alpha_1^2+\beta_1^2]
         -24\delta_1^4\delta_2t(\alpha_0\alpha_1+\beta_0\beta_1) \\
       &
       +48\delta_2[\delta_2(\alpha_0^2+\beta_0^2)+\delta_1(\alpha_0\alpha_1+\beta_0\beta_1)]\big\}\Big\}\bigg\},
\end{cases}
\end{equation*}

and the reduced PDE becomes
\begin{equation*}
   \delta_1^2(P_{\xi\xi}+P_{\eta\eta})+\delta_1^3(P_{\xi}+P_{\eta})
   -4[\delta_1(\alpha_0\alpha_1+\beta_0\beta_1)+\delta_2(\alpha_0^2+\beta_0^2)]P=0.
\end{equation*}

We obtain the solution by the method of separation of variables
\begin{eqnarray}
   P = {\it F_{1}}(\xi){\it F_{2}}(\eta),
\end{eqnarray}
where ${\it F_1}(\xi)$ and ${\it F_2}(\eta)$ is
\begin{equation*}\begin{cases}
   {\it F_1}(\xi) = & C_1e^{\dfrac{\xi}{2}(\sqrt{\delta_1^2+4c_1}-\delta_1)}+C_2e^{-\dfrac{\xi}{2}(\sqrt{\delta_1^2+4c_1}+\delta_1)},\\
   {\it F_2}(\eta) = & C_3e^{\dfrac{\eta}{2\delta_1}(\sqrt{\delta_1^4-4c_1\delta_1^2+16[\delta_1(\alpha_0\alpha_1+\beta_0\beta_1)
                       -\delta_2(\alpha_0^2+\beta_0^2)]}-\delta_1^2)} \\
                     & +C_4e^{-\dfrac{\eta}{2\delta_1}(\sqrt{\delta_1^4-4c_1\delta_1^2+16[\delta_1(\alpha_0\alpha_1+\beta_0\beta_1)
                       -\delta_2(\alpha_0^2+\beta_0^2)]}+\delta_1^2)},
\end{cases}\end{equation*}
where $c_1,\,C_1,\,C_2,\,C_3,\,C_4$ are arbitrary constants.

\begin{itemize}
\item {\bf Case 1.6}
\end{itemize}
\begin{equation*}
\begin{cases}
   M = C(r)+d\theta,\\
   f_1 = f_2=f_3=0,\quad f_4 = -dkt,
\end{cases}
\end{equation*}
where $r=\sqrt{x^2+y^2}$. If $d=0$,
then $C(r)\neq C_0r^{-2}+C_1r^2+c_0$ should be held to make $M$ satisfy the compatibility condition \eqref{condition1}.

We have the similarity variables $\xi, \,\eta, \,P$,
\begin{equation*}
\begin{cases}
   \xi = & x^2+y^2, \\
   \eta = & t, \\
   P = & u e^{d\theta t},
\end{cases}
\end{equation*}

and the reduced PDE becomes
\begin{equation*}
   4\xi P_{\xi\xi}+2\xi(2P_{\xi}-P_{\eta})+(d^2\eta^2-2\xi C(r))P=0.
\end{equation*}

\begin{itemize}
\item {\bf Case 1.8a}
\end{itemize}
\begin{equation*}
\begin{cases}
   M = C(x)+by,\\
   f_1 = f_2=0,\quad k=0,\\
   f_3 = \beta_1t+\beta_0,\quad f_4 =
   \dfrac{b\beta_1}{2}t^2+b\beta_0t,
\end{cases}
\end{equation*}
where $C(x)\neq C_0x^2+ax+c_0$ and $C(x)\neq \frac{C_0}{x^2}+c_0$.

We have the similarity variables $\xi, \,\eta, \,P$,
\begin{equation*}
\begin{cases}
   \xi = & x, \\
   \eta = & t, \\
   P = & u \cdot \exp
   \big\{\dfrac{(\beta_1y+b\beta_1t^2+2b\beta_0t)y}{2(\beta_1t+\beta_0)}\big\},
\end{cases}
\end{equation*}

and the reduced PDE becomes
\begin{equation*}
   4(\beta_1\eta+\beta_0)^2P_{\xi\xi}-8(\beta_1\eta+\beta_0)^2P_{\eta}+b^2\eta^2(\beta_1\eta+2\beta_0)^2P
   -4\beta_1(\beta_1\eta+\beta_0)P-8C(\xi)(\beta_1\eta+\beta_0)^2p=0.
\end{equation*}

We can get the solution by the method of separation of variables
\begin{eqnarray}
   P = {\it F_{1}}(\xi){\it F_{2}}(\eta),\nonumber
\end{eqnarray}
where ${\it F_1}(\varrho),\,{\it F_2}(\theta)$ is the solution of
\begin{equation*}\begin{cases}
   \dfrac{d^2{\it F_{1}}(\xi)}{d\xi^2}+[c_1-2C(\xi)]{\it F_{1}}(\xi)=0,\\
   \dfrac{d{\it F_{2}}(\eta)}{d\eta}+\Big[\dfrac{1}{2}
   -\dfrac{b^2\eta^2(\beta_1\eta+2\beta_0)^2-4\beta_1(\beta_1\eta+\beta_0)}{8(\beta_1\eta+\beta_0)^2}\Big]{\it F_{2}}(\eta)=0.
\end{cases}\end{equation*}
where $c_{1}$ is an arbitrary constant. Given $C(\xi)$, the above ODE systems can be solved directly.

\begin{itemize}
\item {\bf Case 1.8b}
\end{itemize}
\begin{equation*}
\begin{cases}
   M = C(x)+cy^2+by,\\
   f_1 = f_2=0,\quad k=0,\\
   f_3 = \beta_1e^{\sqrt{2c}t}+\beta_2e^{-\sqrt{2c}t},\quad
   f_4 =
   \dfrac{b\beta_1}{\sqrt{2c}}e^{\sqrt{-2c}t}-\dfrac{b\beta_2}{\sqrt{2c}}e^{-\sqrt{2c}t},
\end{cases}
\end{equation*}
where $C(x)\neq C_0x^2+ax+c_0$ and $C(x)\neq \frac{C_0}{x^2}+c_0$.

We have the similarity variables $\xi, \,\eta, \,P$,
\begin{equation*}
\begin{cases}
   \xi = & x, \\
   \eta = & t, \\
   P = & u \cdot \exp \Big\{\dfrac{\beta_1e^{\sqrt{2c}t}-\beta_2e^{-\sqrt{2c}t}}{\beta_1e^{\sqrt{2c}t}+\beta_2e^{-\sqrt{2c}t}}
         \dfrac{y(cy+b)}{\sqrt{2c}}\Big\},
\end{cases}
\end{equation*}

and the reduced PDE becomes
\begin{equation*}
  P_{\xi\xi}-2P_{\eta}-2C(\xi)P=0.
\end{equation*}

We can get the solution by the method of separation of variables
\begin{eqnarray}
   P = {\it F_{1}}(\xi){\it F_{2}}(\eta),\nonumber
\end{eqnarray}
where ${\it F_1}(\varrho),\,{\it F_2}(\theta)$ is the solution of
\begin{equation*}\begin{cases}
   \dfrac{d^2{\it F_{1}}(\xi)}{d\xi^2}+[c_1-2C(\xi)]{\it F_{1}}(\xi)=0,\\
   \dfrac{d{\it F_{2}}(\eta)}{d\eta}+\dfrac{1}{2}{\it F_{2}}(\eta)=0.
\end{cases}\end{equation*}
where $c_{1}$ is an arbitrary constant. Given $C(\xi)$, the above ODE systems can be solved directly.

\end{document}